%% file: sample-sigconf-authordraft.tex
\documentclass[sigconf]{acmart}

\usepackage[utf8]{inputenc} 
\usepackage[T1]{fontenc}    
\usepackage{hyperref}       
\usepackage{url}            
\usepackage{booktabs}       
\usepackage{amsfonts}       
\usepackage{nicefrac}       
\usepackage{microtype}      
\usepackage{xcolor}         
\usepackage{balance}

\usepackage{diagbox}
\usepackage{booktabs}
\usepackage{color}
\usepackage{svg}
\usepackage{subcaption}
\usepackage{graphicx}

\usepackage{booktabs}  
\usepackage{multirow}  
\usepackage{float}    

\begin{document}

\title{FRONTIER-RevRec: A Large-scale Dataset for Reviewer Recommendation}

\author{Qiyao Peng}
\affiliation{%
  \institution{Tianjin University, Tianjin, China}
  \city{}
  \state{}
  \country{}
}
\email{qypeng@tju.edu.cn}

\author{Chen Wang}
\affiliation{%
  \institution{Tianjin University, Tianjin, China}
  \city{}
  \state{}
  \country{}
}
\email{wangchen2023@tju.edu.cn}

\author{Yinghui Wang}
\affiliation{%
  \institution{National Key Laboratory of Information Systems Engineering, Beijing, China}
  \city{}
  \state{}
  \country{}
}
\email{wangyinghui@tju.edu.cn}

\author{Hongtao Liu}
\affiliation{%
  \institution{Du Xiaoman Financial Technology, Beijing, China}
  \city{}
  \state{}
  \country{}
}
\email{htliu@tju.edu.cn}

\author{Xuan Guo}
\authornote{Corresponding author.}
\affiliation{%
  \institution{Tianjin University, Tianjin, China}
  \city{}
  \state{}
  \country{}
}
\email{guoxuan@tju.edu.cn}

\author{Wenjun Wang}
\affiliation{%
  \institution{Tianjin University, Tianjin, China}
  \city{}
  \state{}
  \country{}
}
\email{wjwang@tju.edu.cn}

\input{0-abstract.tex}

\keywords{}

\maketitle

\input{1-introduction.tex}

\input{2-relatedwork.tex}
\input{3-dataset.tex}
\input{4-experiments.tex}
\input{5-conclusion.tex}

\clearpage
\balance
\bibliographystyle{cikm}
\bibliography{cikm}

\end{document}

%% file: 0-abstract.tex
\begin{abstract}
Reviewer recommendation is a critical task for enhancing the efficiency of academic publishing workflows. However, research in this area has been persistently hindered by the lack of high-quality benchmark datasets, which are often limited in scale, disciplinary scope, and comparative analyses of different methodologies.
To address this gap, we introduce FRONTIER-RevRec, a large-scale dataset constructed from authentic peer review records (2007-2025) from the Frontiers open-access publishing platform\footnote{\url{https://www.frontiersin.org/}}. 
The dataset contains 177941 distinct reviewers and 478379 papers across 209 journals spanning multiple disciplines including clinical medicine, biology, psychology, engineering, and social sciences. 
Our comprehensive evaluation on this dataset reveals that content-based methods significantly outperform collaborative filtering. 
This finding is explained by our structural analysis, which uncovers fundamental differences between academic recommendation and commercial domains. 
Notably, approaches leveraging language models are particularly effective at capturing the semantic alignment between a paper's content and a reviewer's expertise.
Furthermore, our experiments identify optimal aggregation strategies to enhance the recommendation pipeline.
FRONTIER-RevRec is intended to serve as a comprehensive benchmark to advance research in reviewer recommendation and facilitate the development of more effective academic peer review systems
\footnote{The FRONTIER-RevRec dataset is available at: \url{https://anonymous.4open.science/r/FRONTIER-RevRec-5D05}}.
\end{abstract}

%% file: 1-introduction.tex
\section{Introduction}

Academic peer review is the primary mechanism for scholarly validation within journals and conferences, which has upheld the standards of scientific communities for centuries \cite{mulligan2013peer, weller2001editorial}. 
With millions of manuscripts submitted annually, editors face mounting challenges in identifying reviewers with relevant expertise \cite{price2017computational, bakker2019peer,nis2016global}. 
Reviewer recommendation systems could effectively alleviate this phenomenon while enhancing peer review quality \cite{zhao2022reviewer, mrowinski2017artificial}, which have been widely userd across different academic platforms \cite{tennant2018state, price2017computational, tennant2017peer}.

Despite its practical importance, research on reviewer recommendation faces following three critical limitations:

\textbf{(1) Limited Dataset Scale}: Existing public datasets for reviewer recommendation suffer from insufficient sample sizes, hampering the training of sophisticated models and robust statistical analysis. The PeerRead dataset \cite{kang2018dataset} contains only approximately 14000 paper submissions, while the OpenReview dataset \cite{gao2019does} includes merely 1542 papers with 1275 reviewers. Zhang et al. \cite{zhang2020multi} constructed the ACM-DL dataset containing 22575 reviewers and 13449 papers with 1944 research field labels. Even the more recent OAG-Bench \cite{zhang2024oag}, while containing over 200000 reviewer-paper pairs, was experimentally validated using only a 40000 instance subset. Limited sample sizes restrict the ability to train complex models and draw statistically robust conclusions about recommendation approaches.

\textbf{(2) Narrow Domain Coverage}: Most available datasets focus almost exclusively on computer science venues, limiting the generalizability of findings to other scientific disciplines. 
PeerRead, OpenReview, and the ACM-DL dataset all center primarily on computer science conferences or repositories like arXiv. 
This disciplinary narrowness fails to capture the diverse domain features across fields such as medicine, biology, psychology, and social sciences. Models trained without exposure to this disciplinary diversity may therefore exhibit limited performance when applied to academic contexts beyond computer science.

\textbf{(3) Insufficient Data Analysis}: Prior research has yet to systematically assess the relative importance of different factors (e.g., textual content/network structure) in building effective reviewer recommendation systems. While studies such as \cite{charlin2013toronto} and \cite{mimno2007expertise} have employed specific techniques like topic modeling or network analysis, there remains a lack of comprehensive comparative evaluation across diverse representation techniques and aggregation strategies using a consistent, large-scale dataset. This gap leaves key questions unresolved: for instance, whether traditional collaborative filtering methods\cite{koren2009matrix, sarwar2001item} are suitable for academic reviewer recommendation, or whether content-based approaches might yield better performance.

In this paper, we introduce FRONTIER-RevRec, a large-scale benchmark dataset constructed from authentic peer review records collected from the Frontiers open-access publishing platform spanning from 2007 to 2025, encompassing 177941 distinct reviewers and their assignments on 478379 papers across 209 journals and 1736 specialized sections. 
This dataset offers cross-disciplinary coverage including clinical medicine, biology \& biochemistry, plant \& animal science, psychology, social sciences, and engineering fields, alleviating both the scale and domain diversity limitations of existing datasets. 
Second, we conduct network analyses to investigate structure differences between academic reviewer recommendation and commercial recommendation, and employ text-based clustering and graph-based topological analyses to understand why traditional collaborate filtering methods might underperform in reviewer recommendation. 
Besides, we examines network connectivity patterns, path characteristics between positive and negative reviewer-paper pairs, and the effectiveness of collaborative signals versus semantic content for modeling reviewer expertise. 
Third, based on the constructed large-scale dataset, we establish comprehensive benchmarks by evaluating diverse recommendation approaches including collaborative filtering methods (LightGCN \cite{he2020lightgcn}, GF-CF \cite{mao2021simplex}), review-based approaches (NARRE \cite{chen2018neural}, DeepCoNN \cite{zheng2017joint}), and pure text-based techniques (TF-IDF, BERT \cite{devlin2019bert}, LLaMA2 \cite{touvron2023llama}).

In all, FRONTIER-RevRec is a comprehensive benchmark dataset for reviewer recommendation and could facilitate future research in this area. The contributions of this paper can be summarized as follows:

\begin{itemize}
    \item We introduce FRONTIER-RevRec, a large-scale, comprehensive dataset for reviewer recommendation that spans multiple disciplines and contains authentic reviewer-paper assignments.
    
    \item We conduct detailed structural analyses that reveal differences between academic reviewer networks and traditional recommendation networks, explaining why collaborative filtering approaches underperform in this reviewer recommendation.
    
    \item We provide benchmarking of diverse recommendation approaches, demonstrating the superiority of content-based methods, particularly those utilizing advanced language models like LLaMA2.
    
    \item We analyze different embedding and aggregation techniques, uncovering that different stages of the recommendation pipeline benefit from different approaches—with averaging excelling at word-to-paper aggregation and sequential models performing best for paper-to-reviewer aggregation.
\end{itemize}

%% file: 2-relatedwork.tex
\section{Related Work}

\subsection{Reviewer Recommendation}

Reviewer recommendation systems aim to match manuscripts with experts who possess relevant expertise from potential candidate pools \cite{price2017computational, balietti2016peer}. This task presents two key challenges: effectively representing text-rich academic papers and accurately modeling reviewer expertise based on their academic activities \cite{charlin2013toronto, zhao2022reviewer}.

Text-based reviewer recommendation methods focused on content matching between manuscripts and reviewers' publication histories \cite{dumais1992automating, mimno2007expertise}. Dumais and Nielsen \cite{dumais1992automating} employed Latent Semantic Indexing to identify similarities between papers and reviewers' CV publications, while Wu et al. \cite{wu2017topic} utilized topic models to represent both papers and reviewers' expertise derived from their publication and review history.
Carpenter et al. \cite{carpenter2025enhancing} proposed a machine learning approach for automated reviewer assignment in astronomical peer review, using LDA topic modeling and optimization algorithms to match proposals with reviewers based on text similarity. The method achieved significant improvements in reviewer-proposal matching accuracy and eliminated manual reassignment efforts at ALMA.

Network-based approaches have leveraged academic network structures for reviewer recommendations \cite{rodriguez2008algorithm, tran2017expert, zhang2013citation,bai2019scientific}. Rodriguez and Bollen \cite{rodriguez2008algorithm} developed particle-swarm algorithms applied to co-authorship networks, while Tran et al. \cite{tran2017expert} utilized expertise networks based on publication records and collaborative relationships. These methods typically embed both papers and reviewers in a shared network space and then compute matching scores.

\subsection{Existing Datasets}

There are only a few public datasets for reviewer recommendation, which are summarized in Table 1. Kang et al. \cite{kang2018dataset} constructed the PeerRead dataset by collecting papers submitted to computer science conferences including ACL, NIPS, and ICLR, along with the reviews they received. It contains approximately 14,000 paper submissions with accept/reject decisions and review text, but lacks explicit reviewer-paper assignment information beyond these specific conferences.

Gao et al. \cite{gao2019does} released the OpenReview dataset, which includes papers and reviews from ICLR conferences. This dataset provides transparent review processes but is limited to a single research community (computer science) and covers only a few years of publications. The dataset contains peer reviews, but primarily focuses on the review content rather than the reviewer assignment process.

Zhang et al. \cite{zhang2020multi} constructed a dataset from the ACM Digital Library containing 22,575 reviewers and 13,449 papers with 1,944 research field labels. Their work transformed paper-reviewer recommendation into multi-label classification using hierarchical representations. However, this dataset remains limited to computer science venues.

Zhang et al. \cite{zhang2024oag} released a reviewer recommendation dataset as part of the OAG-Bench framework, containing over 20,000 reviewer-paper pairs. Their experimental evaluation utilized 40,000 instances from this collection. This dataset represents another contribution to the growing ecosystem of resources for benchmarking reviewer recommendation systems, though the experimental validation was conducted on a subset of the available data.

\begin{table*}
\centering
\caption{Comparison of existing reviewer recommendation datasets.}
\label{tab:datasets}
\setlength{\tabcolsep}{10pt}  
\resizebox{0.9\linewidth}{!}{\begin{tabular}{lccccc}  
\toprule
\textbf{Dataset} & \textbf{Papers} & \textbf{Reviewers} & \textbf{Reviews} & \textbf{Disciplines} & \textbf{Exp. Validation} \\
\midrule
PeerRead  & 14,784 & 10,770 & / & CS & Yes (compre.) \\
OpenReview  & 1,542 & 1,275 & 4,054 & CS & Yes (compre.) \\
ACM-DL dataset  & 13,449 & 22,575 & / & CS & Yes (compre.) \\
OAG-Bench  & 225,478 & 210,069 & / & Multiple & Yes (40K subset) \\
\midrule
\textbf{FRONTIER-RevRec} & \textbf{478,379} & \textbf{177,941} & / & \textbf{Multiple$^*$} & \textbf{Yes (compre.)} \\
\bottomrule
\multicolumn{6}{l}{\footnotesize $^*$209 journals across multiple scientific disciplines.} \\
\end{tabular}}
\end{table*}

In addition, due to data limitations, some researchers have resorted to synthetic datasets or simulations \cite{fiez2020super, stelmakh2021catch} to evaluate algorithms. However, these approaches cannot fully capture the complexities of real-world peer review systems and reviewer behavior.

In summary, most existing public datasets for reviewer recommendation are domain-restricted (primarily computer science), limited in scale, or lack authentic reviewer assignment information across multiple disciplines. Some of them focus on review content rather than the reviewer assignment process, while others don't provide sufficient information to model reviewer expertise over time. Thus, a large-scale cross-disciplinary reviewer recommendation dataset with authentic reviewer assignments is of great value for research in this area.

%% file: 3-dataset.tex
\section{FRONTIER-RevRec Dataset}

\subsection{Dataset Construction}
Research in reviewer recommendation has been consistently hindered by limitations in publicly available datasets, particularly regarding data volume, domain coverage, and access to authentic reviewer assignment processes from academic conferences and journals. To address these constraints and enhance the practical relevance of our study, we constructed FRONTIER-RevRec, a comprehensive dataset for reviewer recommendation extracted from genuine paper review records on the Frontiers open-access publishing platform spanning 2007 to 2025.
The FRONTIER-RevRec dataset encompasses 209 journals with 1,736 specialized sections, yielding a substantial corpus of 518,521 papers. 

\begin{table*}[ht]
\centering
\caption{An example in the FRONTIER-RevRec Dataset.}
\label{tab:metadata_fields}
\begin{tabular}{lp{0.79\textwidth}}
\hline
\textbf{Field} & \textbf{Description} \\
\hline
DOI & 10.3389/fncel.2024.1367476 \\
Authors Names (IDs) & David R. Logan (Null), Jesse Hall (2653031), Laura Bianchi (1301802) \\
Title & A helping hand: roles for accessory cells in the sense of touch across species \\
Abstract & During touch, mechanical forces are converted into electrochemical signals by tactile organs made of neurons, accessory cells, and their shared extracellular spaces.  ......  A greater understanding of touch, which must include a role for accessory cells, is also relevant to emergent technical applications including prosthetics, virtual reality, and robotics.\\
Reviewers Names (IDs) & Simone Pifferi (625657), Federica Mangione (1925325) \\
Domain & Science \\
Journal & Frontiers in Cellular Neuroscience \\
Section & Non-Neuronal Cells \\
Publication Date & 16 February 2024 \\
\hline
\end{tabular}
\end{table*}

We implemented a systematic multi-stage data cleaning process to ensure dataset quality and relevance. Initially, we performed domain filtering, categorizing and retaining only papers classified within five primary domains: Engineering, Health, Humanities and Social Sciences, Science, and Sustainability. This domain-specific approach enabled more targeted analysis of reviewer recommendation patterns across distinct academic fields.

Following domain classification, we conducted metadata completeness assessment where manuscripts with incomplete metadata fields were eliminated, with the exception of author identifiers, which were permitted to be absent in certain cases to maintain dataset comprehensiveness. This selective approach to metadata requirements balanced the need for complete information with preservation of valuable review instances.

For reviewer profile construction, we utilized reviewer identifiers as unique keys to establish comprehensive reviewer profiles based on previously evaluated papers. This approach enabled us to capture reviewer expertise through their documented evaluation history, creating a rich representation of each reviewer's domain knowledge and assessment patterns.

To address potential cold-start issues and ensure sufficient reviewer activity data, we implemented a mitigation strategy whereby reviewers who had assessed fewer than two papers were identified. For these cases, only the specific paper-reviewer interactions were removed rather than eliminating entire paper records. Complete paper records were only deleted when all associated reviewers failed to meet the minimum activity threshold. This nuanced filtering approach preserved valuable manuscript data while ensuring reviewer profiles contained sufficient information for meaningful analysis.

\begin{figure}[t]
    \centering
    \begin{subfigure}[b]{0.48\columnwidth}
        \centering
        \includegraphics[width=\textwidth]{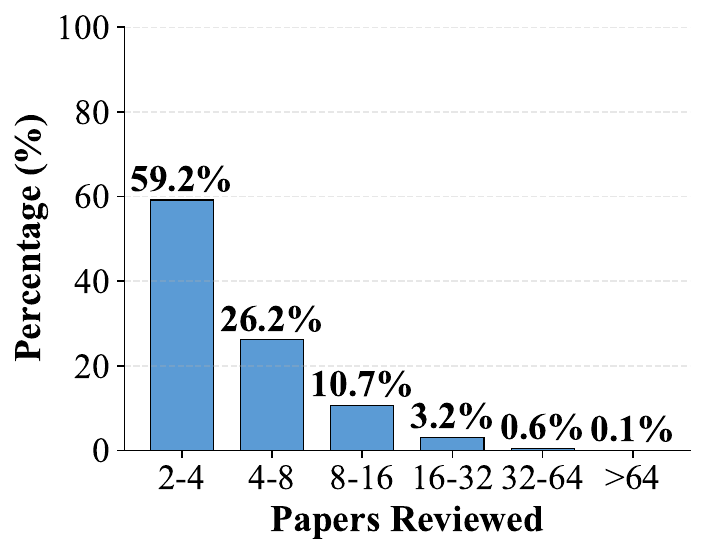}
        \caption{Distribution of Peer Reviewers per Manuscript}
        \label{fig:paper}
    \end{subfigure}
    \hspace{0.02\columnwidth}
    \begin{subfigure}[b]{0.48\columnwidth}
        \centering
        \includegraphics[width=\textwidth]{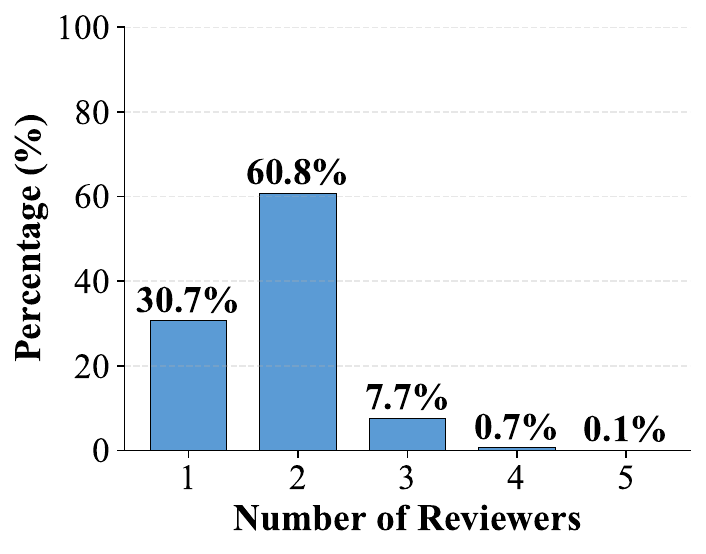}
        \caption{Distribution of Historical Reviewer Workload}
        \label{fig:reviewer}
    \end{subfigure}
    \caption{Analysis of peer review process in FRONTIER-RevRec.}
    \label{fig:combine}
\end{figure}

Finally, we construct a curated dataset comprising 177,941 distinct reviewers and 478,379 papers. The comprehensive statistics of the FRONTIER-RevRec dataset are summarized in Table~\ref{tab:dataset-stats} and the distribution patterns are illustrated in Figure~\ref{fig:combine}.
Each paper contains essential metadata as detailed in Table~\ref{tab:metadata_fields}, including author information, reviewer data, Digital Object Identifiers (DOIs), paper titles, abstracts, domains, journal names, sectional affiliations, and publication dates.
This dataset can provide a consistent basis for matching reviewer expertise with manuscript requirements.

\begin{table}[htbp]
\centering
\caption{FRONTIER-RevRec dataset statistics.}  
\label{tab:dataset-stats}
\resizebox{0.8\linewidth}{!}{\begin{tabular}{lc}
\toprule
\textbf{Statistic} & \textbf{Value} \\
\midrule
Number of Papers & 478,379 \\
Number of Reviewers & 177,941 \\
Average Papers per Reviewer & 4.81 \\
Average Reviewers per Paper & 1.79 \\
Number of Journals & 209 \\
Number of Journal Sections & 1,736 \\
\bottomrule
\end{tabular}}
\end{table}

\subsection{Statistical Characteristic}

\begin{figure}
    \centering
    \includegraphics[width=0.49\textwidth]{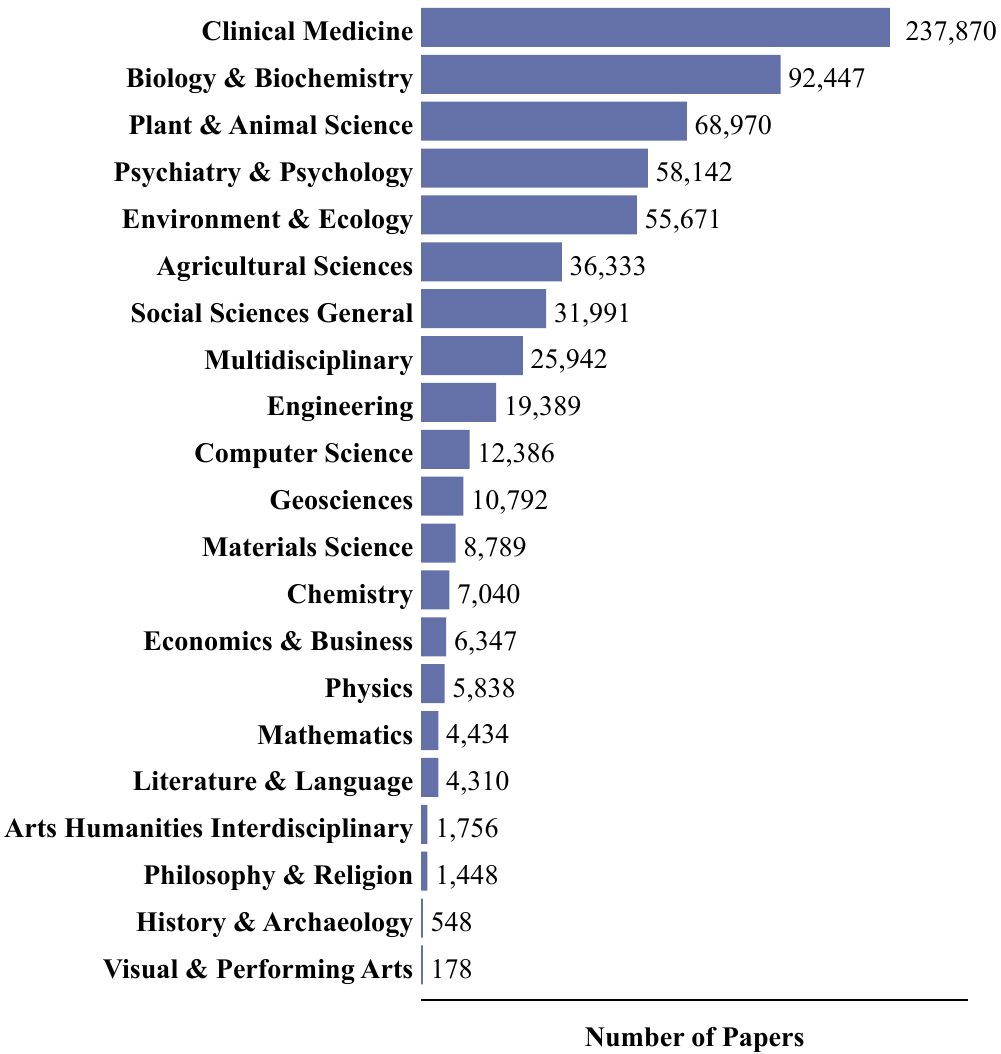}
    \caption{Distribution of Historical Reviewer Workload.}
    \label{fig:distri}
\end{figure}

\subsubsection{Data Distribution and Disciplinary Coverage}

To characterize the scope and balance of FRONTIER-RevRec, we systematically analyzed the distribution of papers across academic disciplines. Given the 1,736 specialized journal sections in the original dataset, we implemented a hierarchical mapping to the Journal Citation Reports' ``Categories by Group'' framework\footnote{\url{https://jcr.clarivate.com/}}, yielding 21 distinct disciplinary groups. This mapping was conducted using Claude 3.7, which enabled nuanced classification where sections could be assigned to primary and secondary disciplinary groups based on their section titles and domain knowledge of academic fields. The prompt-response examples used for this classification process are illustrated in Table~\ref{tab:claude_mapping}.

\begin{table*}[ht]
\centering
\caption{Examples of Prompts and Responses for Disciplinary Mapping Using Claude 3.7.}
\label{tab:claude_mapping}
\begin{tabular}{p{8cm}|p{8cm}}
\hline
\multicolumn{1}{c|}{\textbf{Prompt to Claude 3.7}} & \multicolumn{1}{c}{\textbf{Claude 3.7 Response}} \\
\hline
\textbf{Section: <Non-Neuronal Cells>}, Which of the following category does this section belong to: \newline 1) Agricultural Sciences, 2) Arts \& Humanities, Interdisciplinary, 3) Biology \& Biochemistry, 4) Chemistry, 5) Clinical Medicine, 6) Computer Science, 7) Economics \& Business, 8) Engineering, 9) Environment/Ecology, 10) Geosciences, 11) History \& Archaeology, 12) Literature \& Language, 13) Materials Science, 14) Mathematics, 15) Multidisciplinary, 16) Philosophy \& Religion, 17) Physics, 18) Plant \& Animal Science, 19) Psychiatry/Psychology, 20) Social Sciences, General, 21) Visual \& Performing Arts? \newline\newline Identify the most likely category for this section. Multiple classifications are allowed, but be cautious. Provide the classification(s) separated by commas, and briefly explain your reasoning. & \textbf{3,5} \newline\newline The section ``Non-Neuronal Cells'' primarily belongs to Biology \& Biochemistry (3) and Clinical Medicine (5).\newline\newline Non-neuronal cells (such as glial cells, Schwann cells, etc.) are important components of the nervous system. Research on these cells involves fundamental cellular biology and biochemical processes, which falls within the domain of Biology \& Biochemistry. \newline\newline Additionally, studies on the roles of these cells in various neurological disorders have significant implications for disease diagnosis and treatment, making this topic closely related to Clinical Medicine as well. \\
\hline
\end{tabular}
\end{table*}

As illustrated in Figure \ref{fig:distri}, the resulting disciplinary distribution reveals significant imbalances across research fields. Biomedical and life sciences demonstrate pronounced dominance: Clinical Medicine accounts for 49.7\% (237,870 papers), Biology \& Biochemistry comprises 19.3\% (92,447 papers), and Plant \& Animal Science represents 14.4\% (68,970 papers). Collectively, these three domains constitute over 83\% of the entire dataset. Mid-tier representation includes Psychiatry \& Psychology (12.2\%), Environment \& Ecology (11.6\%), and Agricultural Sciences (7.6\%), while Engineering (4.1\%) and Computer Science (2.6\%) show moderate presence. Notably, humanities and theoretical sciences exhibit minimal representation, with fields such as History \& Archaeology (0.1\%) and Visual \& Performing Arts (0.04\%) comprising less than 1\% of the dataset.

%% file: 4-experiments.tex
\section{Experiments}

\subsection{Content-Based vs. Collaborative Signal Analysis}

To investigate the structural properties of the reviewer recommendation dataset and evaluate the relative importance of different information sources, we conducted both clustering and similarity analyses using text-based and graph-based approaches. This dual methodology is grounded in the historical development of recommendation systems, where both natural language processing (NLP) techniques and graph-based collaborative methods have demonstrated significant effectiveness across various recommendation tasks.

Traditional recommendation systems have successfully leveraged both textual content and collaborative network structures to generate accurate predictions. Content-based methods typically rely on textual similarities between items, while collaborative filtering approaches exploit the underlying graph structure of user-item interactions. In fields such as product recommendation, movie suggestion, and scientific literature discovery, both information sources have proven complementary and mutually reinforcing. Given that our reviewer recommendation dataset inherently contains both rich textual information (paper titles and abstracts) and complex relational structures (reviewer-paper assignments), we sought to determine whether these established paradigms would translate effectively to the specialized domain of reviewer recommendation.

\subsubsection{Content-Based Analysis}

For the text-based approach, we employed the LLaMA model to encode paper titles, generating dense vector representations that effectively capture semantic information. Specifically, embeddings were extracted from the model's final hidden layer, followed by the application of a non-padding average pooling technique to obtain contextually rich representations of each title. The resulting embeddings were subsequently processed using fuzzy c-means clustering algorithms with varying fuzziness parameters ($m$).

Then, we conducted a comprehensive similarity analysis utilizing LLaMA embeddings of paper titles across three distinct levels: global (encompassing all disciplines), intra-category (within the same discipline), and inter-category (between different disciplines). 
Table \ref{tab:similarity_analysis} presents the similarity scores for the overall dataset and two representative disciplinary pairs.

\begin{table}[htbp]
\centering
\small
\caption{LLaMA embedding similarity scores (mean).}
\label{tab:similarity_analysis}
\resizebox{0.9\linewidth}{!}{\begin{tabular}{lccc}
\toprule
\textbf{Category} & \textbf{Global} & \textbf{Intra} & \textbf{Inter} \\
\midrule
All Disciplines & 0.511 & 0.529 & 0.471 \\
Bio/Biochem \& Clinical Med & 0.548 & 0.564 & 0.548 \\
Geosciences \& Materials Sci & 0.473 & 0.494 & 0.450 \\
\bottomrule
\end{tabular}}
\end{table}

The data presented in the table demonstrates that intra-category similarity consistently surpasses global similarity, which in turn exceeds inter-category similarity. This pattern confirms that text embeddings effectively capture disciplinary distinctions. 
Besides, the Geosciences and Materials Science pairing exhibits pronounced differentiation between categories, whereas the Biology/Biochemistry and Clinical Medicine pairing reveals more subtle differences between global and inter-category similarity, reflecting the substantial topical overlap between these closely related disciplines. These findings provide valuable insights into the semantic structure of academic literature and its potential applications in reviewer recommendation systems.

\subsubsection{Graph-Based Analysis}

We construct a reviewer-paper bipartite graph representing assignment relationships between reviewers and papers. From this bipartite structure, we derived a paper-paper relationship matrix based on shared reviewers, effectively transforming indirect connections through common reviewers into direct paper-to-paper relationships. This relational structure was analyzed using a hybrid methodology combining spectral clustering and fuzzy c-means, with the resulting spectral embeddings subjected to fuzzy c-means clustering to identify overlapping communities.

\begin{figure}[htbp]
    \centering
    \includegraphics[width=0.98\columnwidth]{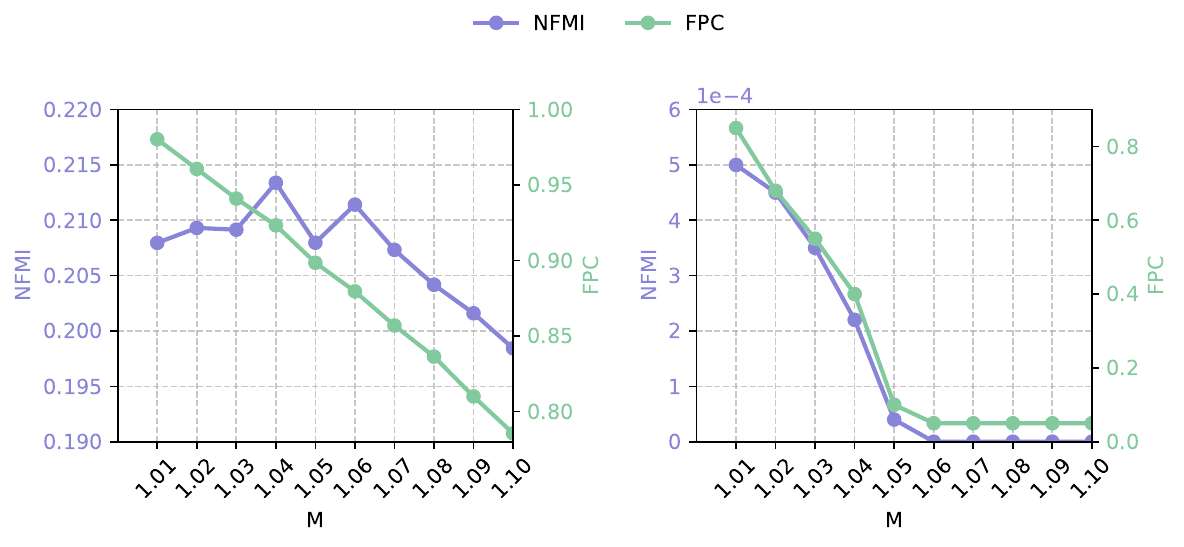}
    
    \parbox{0.49\columnwidth}{\centering (a) Text-based clustering using LLaMA embeddings of paper titles}
    \hfill
    \parbox{0.49\columnwidth}{\centering (b) Graph-based clustering using paper-reviewer network topology}
    
    \caption{Comparison of clustering quality metrics across varying fuzziness parameters ($m$).}
    \label{fig:combined}
\end{figure}

Figure \ref{fig:combined} illustrates the relationship between the fuzziness parameter and two key clustering quality metrics (NFMI and FPC) for both text-based and graph-based approaches. While the text-based method exhibits a non-monotonic NFMI pattern with a peak value of approximately 0.214 at $m = 1.04$, the graph-based clustering method displays markedly different characteristics. Both NFMI and FPC metrics demonstrate a rapid monotonic decrease as the fuzziness parameter increases, with values approaching zero beyond $m = 1.05$. 

This sharp decline indicates that collaborative signals based on shared reviewers are substantially weaker in this dataset compared to the semantic signals captured by the text-based approach. This finding diverges significantly from conventional recommendation domains, where collaborative signals typically provide substantial predictive power. The observed discrepancy highlights the unique challenges of academic reviewer recommendation, where the specialized nature of expertise and sparse reviewer-paper assignments limit the effectiveness of traditional collaborative filtering approaches.

In all, our comparative analysis reveals a critical insight: textual content provides significantly more meaningful structural information than collaborative network signals in reviewer recommendation contexts. Text-based approaches effectively reveal natural disciplinary boundaries with appropriate sensitivity to interdisciplinary overlap, while graph-based approaches fail to identify robust community structures. These findings suggest that effective reviewer recommendation systems should prioritize text-based features capturing semantic relationships rather than relying heavily on collaborative filtering approaches leveraging network topology.

\subsubsection{Compared with traditional recommendation datasets}

To investigate the characteristics distinguishing academic reviewer networks from commercial recommendation networks, we analysis the topology of our dataset and commercial benchmarks (Amazon, Gowalla, Yelp), focusing on key network properties including structural properties, and node reachability.
Table~\ref{tab:dataset_comparison} presents the key network characteristics of these datasets.

The structural analysis revealed topological differences between academic and commercial networks: (1) the Frontiers network exhibited extreme fragmentation with 6,137 connected components, contrasting sharply with the single connected component characteristic of all commercial datasets; (2) While the largest connected component in the Frontiers dataset contained 96.5\% of total nodes, its estimated diameter of 48 hops significantly exceeded the 7-8 hop diameters observed in commercial networks.

Reachability metrics provided insights into network connectivity patterns: (1) in the Frontiers dataset, only 5\% of both positive and negative test pairs achieved reachability within 30 hops, with average path lengths exceeding 25 hops; (2) commercial recommendation datasets, by contrast, demonstrated 100\% reachability for all test pairs, with concise average path lengths of 2.01-2.06 hops for positive pairs and 2.57-2.72 hops for negative pairs.

The absence of discriminative path characteristics between positive and negative pairs in the Frontiers dataset explains why collaborative filtering performs poorly in academic contexts. While commercial networks exhibit clear signals where positive pairs consistently have shorter paths than negative pairs, academic reviewer networks lack this topological distinction necessary for identifying appropriate assignments. These structural differences limit the effectiveness of traditional collaborative filtering approaches in academic peer review domains.

\begin{table*}
\centering
\caption{Comparison of network properties across recommendation datasets.}
\label{tab:dataset_comparison}
\resizebox{0.88\linewidth}{!}{\begin{tabular}{llrrrr}
\toprule
\textbf{Category} & \textbf{Property} & \textbf{Frontiers} & \textbf{Amazon} & \textbf{Gowalla} & \textbf{Yelp} \\
\midrule
Network Structure & Total nodes & 656,320 & 144,242 & 70,839 & 69,716 \\
 & Connected components & 6,137 & 1 & 1 & 1 \\
 & Largest component \% & 96.5\% & 100\% & 100\% & 100\% \\
 & Network diameter & 48 & 7 & 8 & 7 \\
\midrule
Positive Pairs & Average path length & 25.26 (25.42$^*$) & 2.06 & 2.03 & 2.01 \\
 & Reachability rate & 5.15\% (4.24\%$^*$) & 100\% & 100\% & 100\% \\
\midrule
Negative Pairs& Average path length & 25.24 (25.36$^*$) & 2.57 & 2.72 & 2.60 \\
 & Reachability rate & 5.34\% (4.47\%$^*$) & 100\% & 100\% & 100\% \\
\bottomrule
\multicolumn{6}{l}{\footnotesize $^*$Values in parentheses are for the main connected component only.} \\
\end{tabular}}
\end{table*}

\subsection{Benchmark Experiments on Reviewer Recommendation}

\subsubsection{Experimental Setup}

For evaluation, we employ a leave-one-out strategy: one paper per reviewer was randomly selected for the test set, while the remaining papers constituted the training set. Table \ref{tab:data-split} presents the statistics of our data partitioning.

\begin{table}[htbp]
\centering
\caption{Statistics of training and test sets.}
\label{tab:data-split}
\resizebox{0.8\linewidth}{!}{\begin{tabular}{lcc}
\toprule
\textbf{Statistic} & \textbf{Training Set} & \textbf{Test Set} \\
\midrule
Number of Reviewers & 177,941 & 177,941 \\
Number of Papers & 430,435 & 159,713 \\
\bottomrule
\end{tabular}}
\end{table}

In the training phase, we applied negative sampling techniques where each positive reviewer-paper pair was matched with four randomly selected negative examples. During testing, each paper was associated with a candidate set of 25 reviewers, comprising the ground truth reviewer (positive example) and 24 randomly selected negative examples from reviewers who had not previously assessed the paper.

\subsubsection{Problem Definition} The reviewer recommendation task aims to identify the most suitable reviewers for a given target paper $p$ from a candidate reviewer set $R$. This task can be formalized as learning a ranking function: $f: (p, R) \rightarrow L$, where $L$ represents the relevance ranking of reviewers with respect to paper $p$. During evaluation, reviewers who are actually assigned to the target paper are considered as the ground truth.

\subsubsection{Compared Methods}
Based on our analysis revealing the relative importance of collaborative signals versus textual content, we systematically evaluate methods across three categories that represent different information utilization strategies in reviewer recommendation systems.

\begin{itemize}
    \item \textbf{Collaborative Signal-based Methods}. Graph-based collaborative filtering approaches including \textbf{LightGCN} \cite{he2020lightgcn} and \textbf{GF-CF} \cite{mao2021simplex} rely solely on reviewer-paper interaction patterns without considering textual content. For both methods, we construct a bipartite graph from paper-reviewer interactions: \textbf{LightGCN} simplifies Graph Convolutional Networks by removing nonlinearities and learns node embeddings through message passing, while \textbf{GF-CF} leverages high-order connectivity patterns in the same bipartite graph structure.
    \item \textbf{Review-based Recommendation Methods}. We adapt existing review-based recommendation frameworks to utilize textual information in the peer review domain. Both methods use paper titles as textual input: \textbf{NARRE} \cite{chen2018neural} incorporates textual content to enhance recommendation accuracy by treating paper titles as review text and learning reviewer embeddings from historical behavior, while \textbf{DeepCoNN} \cite{zheng2017joint} employs a dual-network architecture to jointly model papers and reviewers, computing matching scores via factorization machine.
    \item \textbf{Pure Text-based Methods}. We explore a spectrum of text representation techniques, ranging from traditional statistical methods to advanced neural language models. As shown in Figure~\ref{fig:Matching_Framework}, we maintain a consistent construction process: (1) we use text encoder (including \textbf{TF-IDF}, \textbf{BERT-base-uncased} \cite{devlin2019bert}, and \textbf{LLaMA2-7B} \cite{touvron2023llama} respectively) to encode paper embedding; (2) we use the aggregation mechanism (e.g., mean pooling) to encode reviewer representations based on previously reviewed paper representations. The final matching scores are  calculated based on target paper embeddings and reviewer representations. For cold-start reviewers, we utilize ID embeddings as initial representations.
\end{itemize}

\subsubsection{Evaluation Metrics}

We evaluated performance using standard information retrieval metrics at different cutoff thresholds $k \in \{1, 5, 10, 15\}$:

\begin{itemize}
\item \textbf{Precision@k (P@k)}: Measures the proportion of recommended reviewers that are relevant:
\begin{equation}
\text{P@k} = \frac{1}{|D|} \sum_{d \in D} \frac{|R_d \cap S_k(d)|}{k}
\end{equation}
where $R_d$ represents the set of relevant reviewers for paper $d$, and $S_k(d)$ represents the top-$k$ recommended reviewers.

\item \textbf{Recall@k (R@k)}: Measures the proportion of relevant reviewers that are successfully recommended:
\begin{equation}
\text{R@k} = \frac{1}{|D|} \sum_{d \in D} \frac{|R_d \cap S_k(d)|}{|R_d|}
\end{equation}

\item \textbf{Normalized Discounted Cumulative Gain@k (NDCG@k)}: Evaluates the ranking quality of recommended reviewers, considering both relevance and position:
\begin{equation}
\text{NDCG@k} = \frac{1}{|D|} \sum_{d \in D} \frac{\text{DCG}_k(d)}{\text{IDCG}_k(d)}
\end{equation}
where $\text{DCG}_k(d) = \sum_{i=1}^{k} \frac{2^{rel_i} - 1}{\log_2(i+1)}$ and $\text{IDCG}_k(d)$ is the ideal DCG value obtained by sorting reviewers by their true relevance.
\end{itemize}

In these equations, $D$ represents the set of test papers, and $\{\text{recommended reviewers}\}_k$ denotes the top-$k$ reviewers recommended by a model.

\subsubsection{Experimental Results}

Table~\ref{tab:benchmark_results} presents the benchmark results across all methods and metrics, we can find:

(1) LLaMA2 with fine-tuning demonstrates remarkable performance, achieving the highest scores across all metrics. BERT similarly shows strong performance, though consistently below LLaMA2. This performance differential can be attributed to LLaMA2's larger model capacity and more extensive pretraining on large scale corpus, enabling it to better capture the nuanced semantic in textual contents.

(2) The text-aware recommendation methods occupy the middle tier in performance rankings. NARRE achieves particularly notable results, outperforming DeepCoNN across all metrics. This superiority likely stems from NARRE's attention mechanism, which effectively focuses on more relevant aspects of paper titles when matching with reviewer.

(3) The graph-based methods demonstrate the weakest performance overall. These results align with our earlier structural analysis that revealed the limitations of collaborative signals in academic reviewer networks. The minimal differences in network path characteristics between positive and negative reviewer-paper pairs significantly constrain the discriminative power of graph-based methods.

In all, the superior results of language model-based approaches, particularly LLaMA2, highlight the importance of capturing deep semantic relationships in academic content for accurate reviewer-paper matching.

\subsubsection{Impact of Candidate Size}

\begin{figure}[htbp]
    \centering
    \includegraphics[width=\columnwidth]{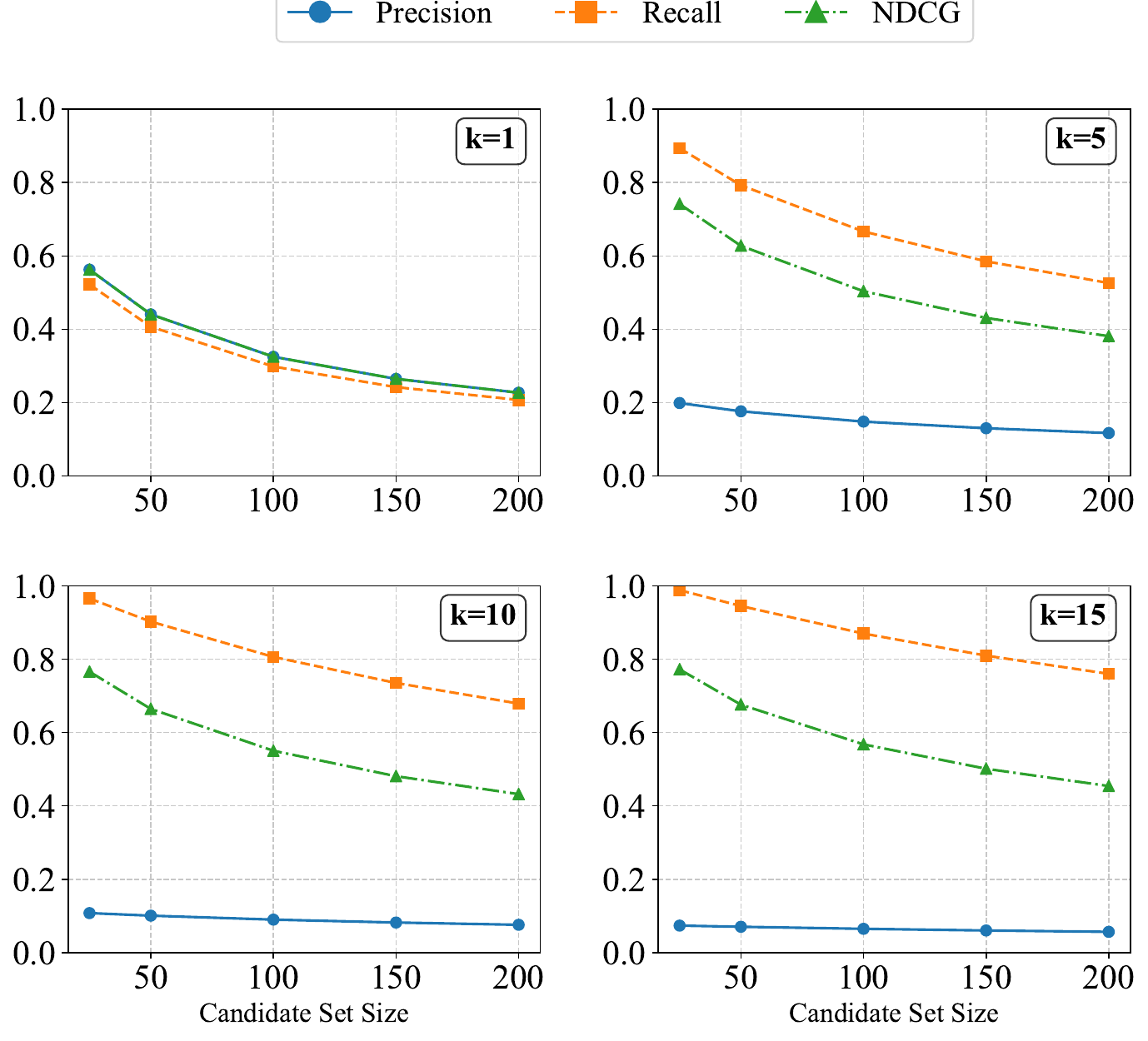}    
    \caption{Performance of LLaMA2 with different candidate set sizes for reviewer recommendation.}
    \label{fig:candidate_sizes}
    
\end{figure}

Considering the candidate size could impact the recommendation performance, we conduct experiments to analysis the impact of candidate set size on LLaMA2 performance. 
Figure~\ref{fig:candidate_sizes} presents results with candidate set sizes ranging from 25 to 200 reviewers.
As expected, performance decreases as the candidate set size increases, with precision at top-1 (P@1) dropping from 0.5627 with 25 candidates to 0.2270 with 200 candidates. This decline is natural, as the task becomes more challenging with larger candidate pools. However, even with 50 candidates, LLaMA2 maintains a P@1 of 0.4406. This demonstrates it's robust discriminative power across varying levels of task difficulty.
Besides, the relative performance patterns observed in our main benchmark remain consistent across different candidate set sizes. Text-based methods maintain their advantage over graph-based approaches regardless of candidate pool size. This consistency demonstrates our conclusion that textual content provides more meaningful signals for reviewer recommendation than network structure.

\begin{table*}
\centering  
\caption{Performance comparison of different reviewer recommendation methods on the Frontiers dataset.}
\label{tab:benchmark_results}
\large  
\setlength{\tabcolsep}{4pt}  
\begin{tabular}{lcccccccccccc}  
\toprule
\multirow{2}{*}{\textbf{Method}} & \multicolumn{3}{c}{\textbf{@1}} & \multicolumn{3}{c}{\textbf{@5}} & \multicolumn{3}{c}{\textbf{@10}} & \multicolumn{3}{c}{\textbf{@15}} \\
\cmidrule(lr){2-4} \cmidrule(lr){5-7} \cmidrule(lr){8-10} \cmidrule(lr){11-13}
 & P & R & NDCG & P & R & NDCG & P & R & NDCG & P & R & NDCG \\
\midrule
LightGCN & 0.0755 & 0.0695 & 0.0755 & 0.0494 & 0.2226 & 0.1475 & 0.0462 & 0.4146 & 0.2105 & 0.0455 & 0.6123 & 0.2639 \\
GF-CF & 0.1155 & 0.1087 & 0.1155 & 0.0761 & 0.3520 & 0.2333 & 0.0631 & 0.5777 & 0.3075 & 0.0554 & 0.7555 & 0.3557 \\
NARRE & 0.2519 & 0.2308 & 0.2519 & 0.1542 & 0.6949 & 0.4307 & 0.0997 & 0.8957 & 0.548 & 0.0719 & 0.9679 & 0.5677 \\
DeepConn & 0.1912 & 0.1749 & 0.1912 & 0.1336 & 0.6020 & 0.4003 & 0.0939 & 0.8434 & 0.4808 & 0.0705 & 0.9493 & 0.5098 \\
TF-IDF & 0.1743 & 0.1580 & 0.1743 & 0.0903 & 0.4054 & 0.2896 & 0.0668 & 0.5999 & 0.3537 & 0.0554 & 0.7455 & 0.3932 \\
BERT & 0.4886 & 0.4528 & 0.4886 & 0.1930 & 0.6685 & 0.6930 & 0.1068 & 0.9596 & 0.7237 & 0.0733 & 0.9869 & 0.7313 \\
LLaMA2 & \textbf{0.5627} & \textbf{0.5220} & \textbf{0.5627} & \textbf{0.1988} & \textbf{0.8940} & \textbf{0.7417} & \textbf{0.1075} & \textbf{0.9656} & \textbf{0.7660} & \textbf{0.0734} & \textbf{0.9884} & \textbf{0.7722} \\
\bottomrule
\multicolumn{13}{l}{\footnotesize * Bold values indicate the best results.} \\
\end{tabular}
\end{table*}

\begin{figure}[htbp]
    \centering
    \includegraphics[width=0.6\columnwidth]{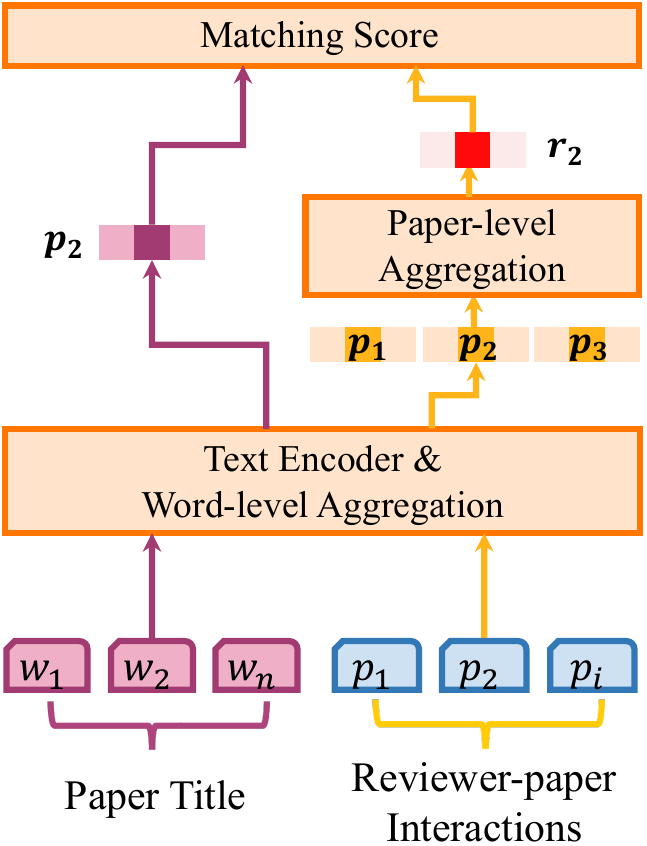}
    \caption{Paper-Reviewer Matching Framework Using Text Encoding.}
    \label{fig:Matching_Framework}
\end{figure}

\begin{figure}[htbp]
    \centering
    \includegraphics[width=\columnwidth]{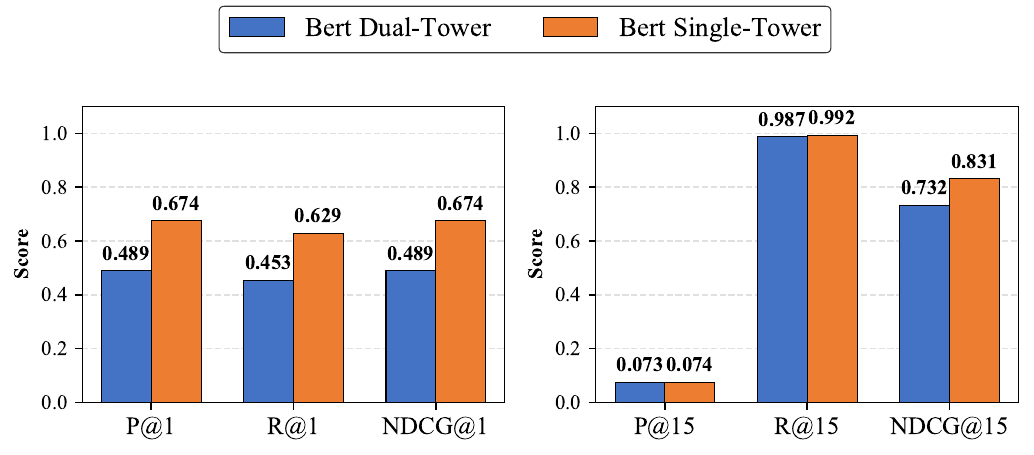}
    \caption{Performance comparison of Dual-Tower and Single-Tower models.}
    \label{fig:bert-comparison}
\end{figure}

\subsubsection{Analysis of Dual-/Single Tower}

In this study, we compare single-tower and dual-tower architectures for reviewer recommendation. The single-tower model directly concatenates paper titles with reviewers' historical review records as input to a single encoder, which then computes matching scores end-to-end. In contrast, the dual-tower architecture processes paper information and reviewer profiles through separate encoders before computing matching scores between the resulting representations. Experimental results (as shown in Figure~\ref{fig:bert-comparison}) demonstrate that the single-tower model consistently outperforms the dual-tower approach across all evaluation metrics. This superior performance can be attributed to the single-tower's ability to learn interactive features between papers and reviewers during the encoding process, enabling more sophisticated cross-modal understanding compared to the dual-tower's separate encoding followed by simple matching. However, the single-tower architecture faces computational efficiency challenges as it must process each paper-reviewer pair individually during inference.

\subsubsection{Analysis of Word-/Paper-level Embedding Aggregation Mechanism}

To  evaluate the impact of different aggregation mechanism on recommendation performance, we conduct experiments using BERT as our consistent base embedding model. 
As shown in Figure~\ref{fig:Matching_Framework}, our investigation focused on two critical aggregation levels: (1) Word-level aggregation: aggregating word embeddings to paper representations; (2) Paper-level aggregation: learning reviewer representations via aggregating paper embeddings.

\begin{figure}[htbp]
    \centering
    \includegraphics[width=\columnwidth]{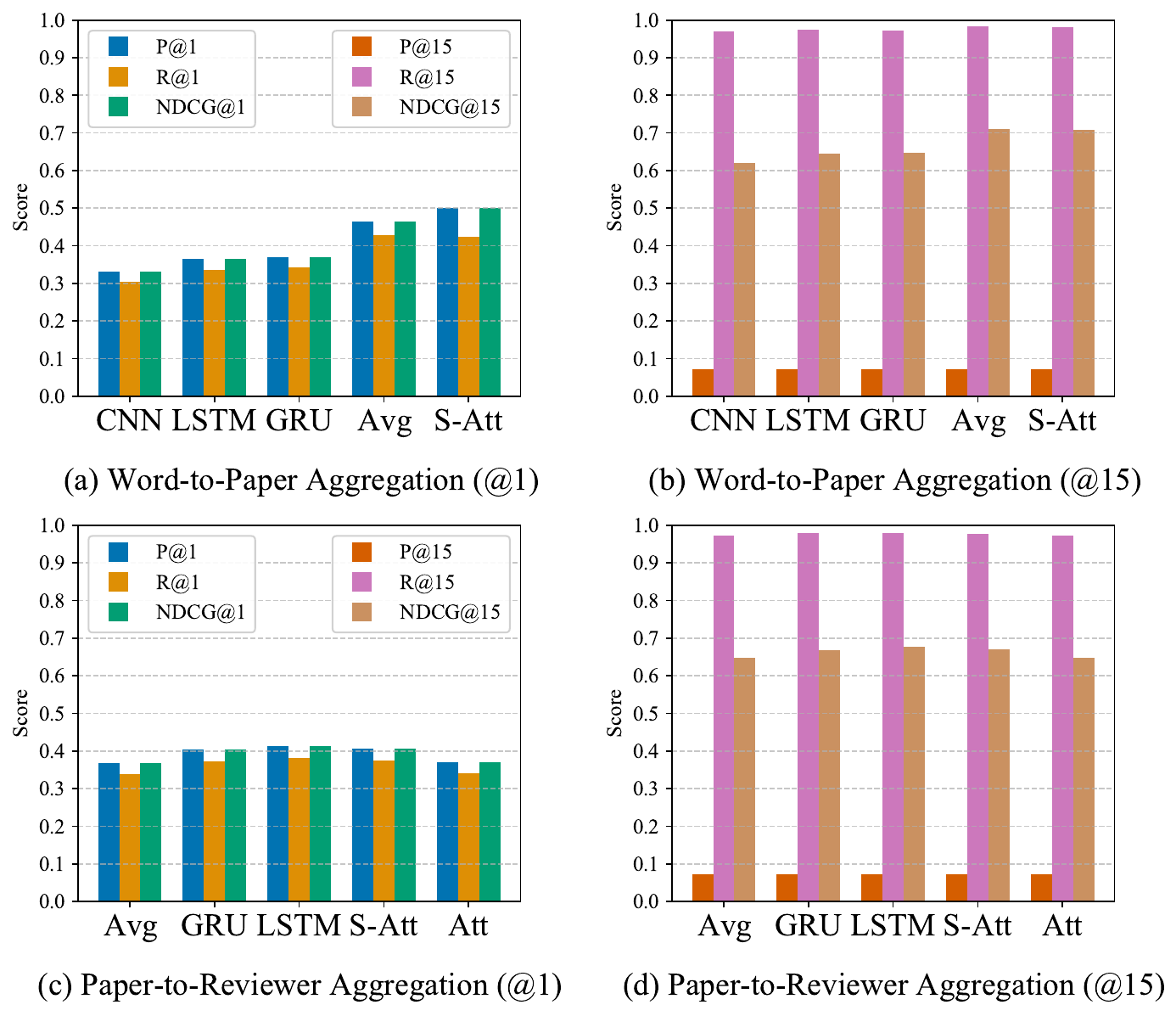}    
    \caption{Performance comparison of different aggregation methods for word-to-paper and paper-to-reviewer representations using BERT.}
    \label{fig:bert_aggregation}
    
\end{figure}

\paragraph{Word-to-Paper Aggregation Methods}

For transforming word-level BERT embeddings into paper representations, we implemented five approaches:

\begin{itemize}
\item \textbf{CNN}: Apply multiple convolutional filters of varying widths to the sequence of word embeddings, capturing local n-gram patterns in the paper titles, followed by max-pooling to obtain fixed-length representations.
\item \textbf{LSTM}: Process the sequence of word embeddings using a Long Short-Term Memory network, capturing dependencies between words while maintaining the sequential nature of language.
\item \textbf{GRU}: Employ a Gated Recurrent Unit, a simplified variant of LSTM, to sequentially process word embeddings while addressing vanishing gradient issues.
\item \textbf{Average}: Simply averages all word embeddings in the title, treating each word as equally important to the overall semantic meaning of the paper.
\item \textbf{Self-Attention}: Implement a multi-head self-attention mechanism where each word can attend to all other words in the title, dynamically weighting their importance based on contextual relevance.
\end{itemize}

\paragraph{Paper-to-Reviewer Aggregation Methods}
For aggregating paper embeddings into reviewer expertise profiles, we used the following approaches:

\begin{itemize}
\item \textbf{Average}: Compute the element-wise mean of all paper embeddings in a reviewer's history, creating a centroid that represents their overall expertise.
\item \textbf{GRU}: Treat a reviewer's paper history as a sequence and processes it using a Gated Recurrent Unit, potentially capturing the evolution of research interests over time.
\item \textbf{LSTM}: Similar to GRU but using Long Short-Term Memory networks to model longer-range dependencies in a reviewer's publication history.
\item \textbf{Self-Attention}: Employ a multi-head self-attention mechanism where each paper in a reviewer's history can attend to all other papers, identifying complex relationships between previously reviewed papers.
\item \textbf{Attention}: Implement a simple attention mechanism that computes a weighted sum of paper embeddings in a reviewer's history, with weights determined by similarity to the target paper being considered for review.
\end{itemize}

Figure~\ref{fig:bert_aggregation} presents the performance of different aggregation methods. We can find that:

(1) For word-to-paper aggregation, we observe that simpler methods outperform more complex sequential models. Average pooling achieves the best performance, closely followed by self-attention. Both substantially outperform the sequential models: GRU or LSTM. This suggests that for academic paper titles, which are typically concise and information-dense, sophisticated sequential processing may be less beneficial than approaches that treat each word as equally important (averaging) or that dynamically focus on the relevant terms (self-attention).

(2) In contrast, when aggregating paper embeddings to construct reviewer representations, the LSTM achieves the best performance, while simple averaging  and basic attention lag behind.
This is because more recent papers potentially are important than earlier ones for modeling reviewers. 
The LSTM can capture this temporal aspect which could reflect how reviewers' interests and expertise evolve over time.

(3) Another interesting observation is that the performance gap between the best and worst aggregation methods is larger for word-to-paper aggregation (39.6\% difference between average pooling and CNN) than for paper-to-reviewer aggregation (12.5\% difference between LSTM and average pooling). This indicates that the choice of aggregation method may be more critical at the word-to-paper level than at the paper-to-reviewer level.

These findings have important implications for the design of reviewer recommendation systems. They highlight that different levels of aggregation in the recommendation pipeline—from words to paper representations, and from papers to reviewer profiles—benefit from different strategies. Specifically, simpler methods like averaging and self-attention work better for aggregating words into paper representations, while sequential models like LSTM are more effective for aggregating papers into reviewer profiles.

%% file: 5-conclusion.tex
\section{Conclusion}

FRONTIER-RevRec provides a large-scale, cross-disciplinary dataset for reviewer recommendation research with 177,941 reviewers and 478,379 papers across 209 journals. Our analysis shows distinct structural patterns in academic reviewer networks compared to commercial networks, evidenced by fragmentation patterns and path characteristics. In experiments, text-based methods using language models significantly outperform collaborative filtering approaches, achieving superior precision and recall across metrics. Besides, we observe that different aggregation techniques perform optimally at different pipeline stages—averaging excels for word-to-paper aggregation while sequential models better capture reviewer expertise evolution. This dataset establishes empirical benchmarks for academic reviewer recommendation systems and enables further research in academic data analysis. In the future, we will: (1) develop cross-domain recommendation model to address disciplinary imbalances and reviewer scarcity in emerging fields; (2) design fairness-aware recommendation algorithms that ensure equitable reviewer distribution across demographics, institutions, and geographic regions while maintaining quality standards. These future directions will contribute to more effective, fair, and scalable academic peer review systems.